\documentclass[citeautoscript,reprint,twocolumn,amsmath,amssymb,aip,jcp,floatfix]{revtex4-1}
\usepackage{txfonts}

\usepackage{hyperref}
\hypersetup{
  colorlinks=true,
  linkcolor=blue,
  citecolor=red,
  linktoc=all,
  pdfauthor=B. Mussard,
  pdftitle=Spin-unrestricted random-phase approximation with range separation: Benchmark on atomization energies and reaction barrier heights,
  pdfdisplaydoctitle=true,
  pdfpagelayout=SinglePage,
  pdfstartview=Fit,
  pdfstartpage=1,
  bookmarksopen=false
}

\newcommand\equ[2]{\begin{align}\label{#1} #2 \end{align}}

\usepackage{graphicx}
\usepackage{epstopdf}
\usepackage[export]{adjustbox}
\graphicspath{{data_plot/}}
\newcommand\fig[7]{
 \includegraphics[width=#7,height=#6,keepaspectratio=true,trim=#2 #3 #4 #5, clip=true]{#1}
}

\usepackage{tabularx}
\newcolumntype{R}{>{\raggedleft\arraybackslash}X}
\newcolumntype{C}{>{\centering\arraybackslash}X}

\usepackage[T1]{fontenc}

\everymath{\displaystyle}

\newcommand{\bra}[1]{\ensuremath{\langle #1 \vert}}
\newcommand{\ket}[1]{\ensuremath{\vert #1  \rangle}}
\newcommand{\braket}[2]{\ensuremath{\langle  #1 \vert #2  \rangle}}
\renewcommand{\d}{\ensuremath{\text{d}}}
\newcommand{\I}{\ensuremath{\text{I}}}
\newcommand{\II}{\ensuremath{\text{II}}}
\newcommand{\dRPA}{\ensuremath{\text{dRPA}}}
\newcommand{\RPAx}{\ensuremath{\text{RPAx}}}
\newcommand{\tr}{\ensuremath{\text{tr}}}
\renewcommand{\c}{\ensuremath{\text{c}}}

\usepackage{bm}

\renewcommand\b[1]       {\mathbf{#1}}

\newcommand\FloatRef[1]  {\ref{#1}}

\newcommand\figCOMPARE{
\begin{figure*}[!htb]
\centering
\begin{minipage}[t]{.48\linewidth}

  \begin{minipage}[t]{.95\linewidth}
   \fig{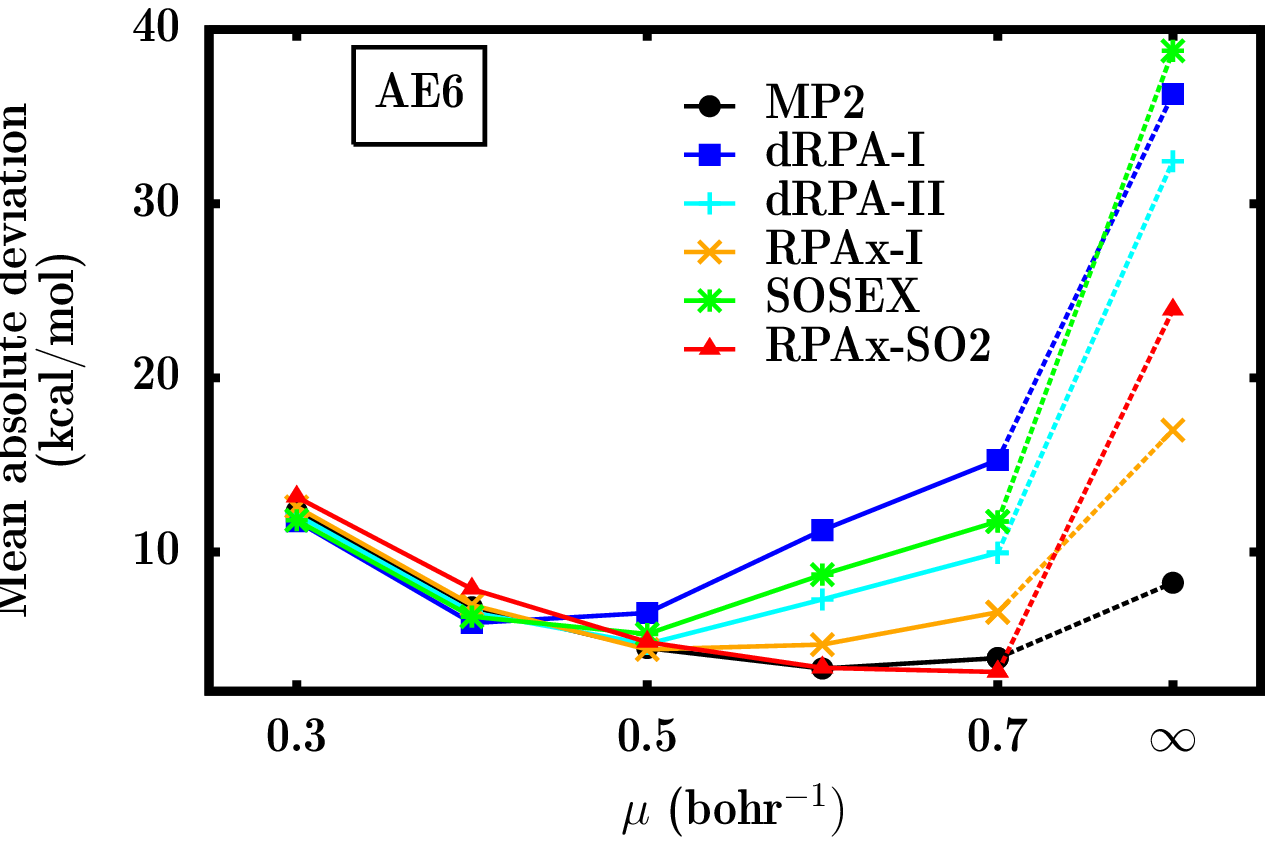}{2cm}{0cm}{0cm}{0cm}{8cm}{\linewidth}
   \label{fig:depMUAE6}
  \end{minipage}%
  
  \begin{minipage}[t]{.95\linewidth}
   \fig{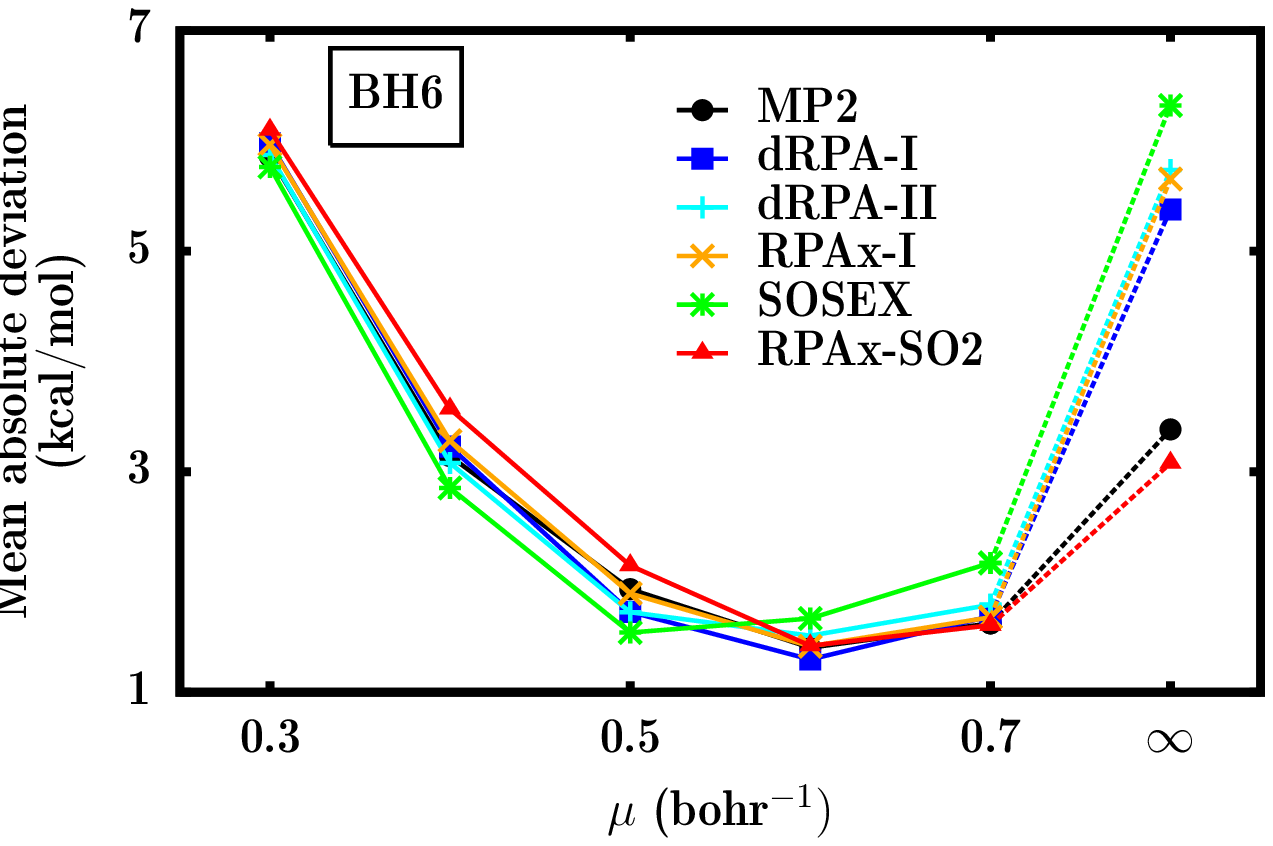}{2cm}{0cm}{0cm}{0cm}{8cm}{\linewidth}
   \label{fig:depMUBH6}
  \end{minipage}

  \caption{Mean absolute deviations for the AE6 and BH6 datasets as functions of the range-separation parameter $\mu$
  for range-separated calculations using the cc-pVQZ basis set, the srPBE functional and different post-RSH long-range correlation methods
  (MP2, dRPA-I, dRPA-II, RPAx-I, SOSEX, and RPAx-SO2). The case $\mu=\infty$ corresponds to full-range post-HF calculations.
  The reference values are the non-relativistic FC-CCSD(T)/cc-pVQZ-F12 values of Refs.~\onlinecite{Haunschild:12a,Haunschild:13}.
  }
  \label{fig:depMU}
\end{minipage}\qquad%
\begin{minipage}[t]{.48\linewidth}

  \begin{minipage}[t]{.95\linewidth}
   \fig{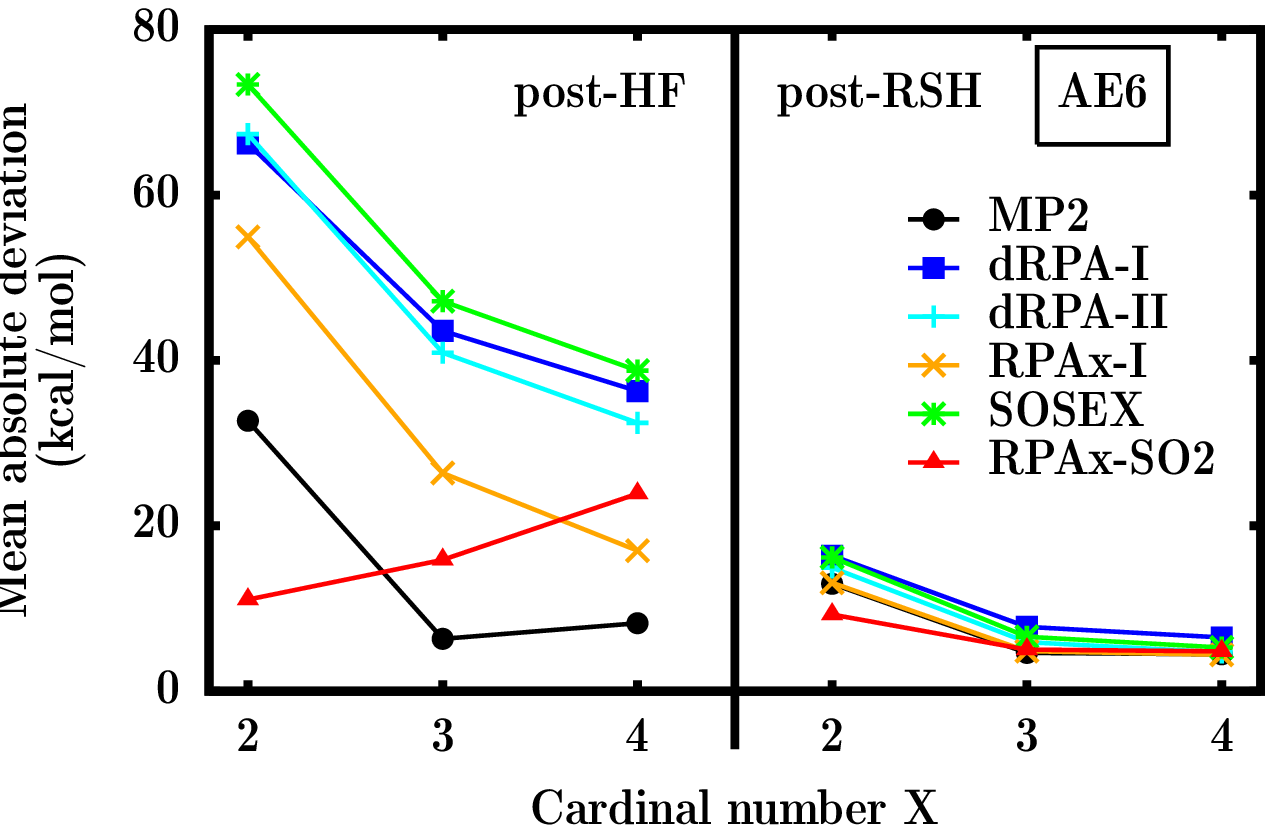}{2cm}{0cm}{0cm}{0cm}{8cm}{\linewidth}
   \label{fig:depBASISAE6}
  \end{minipage}%
  
  \begin{minipage}[t]{.95\linewidth}
   \fig{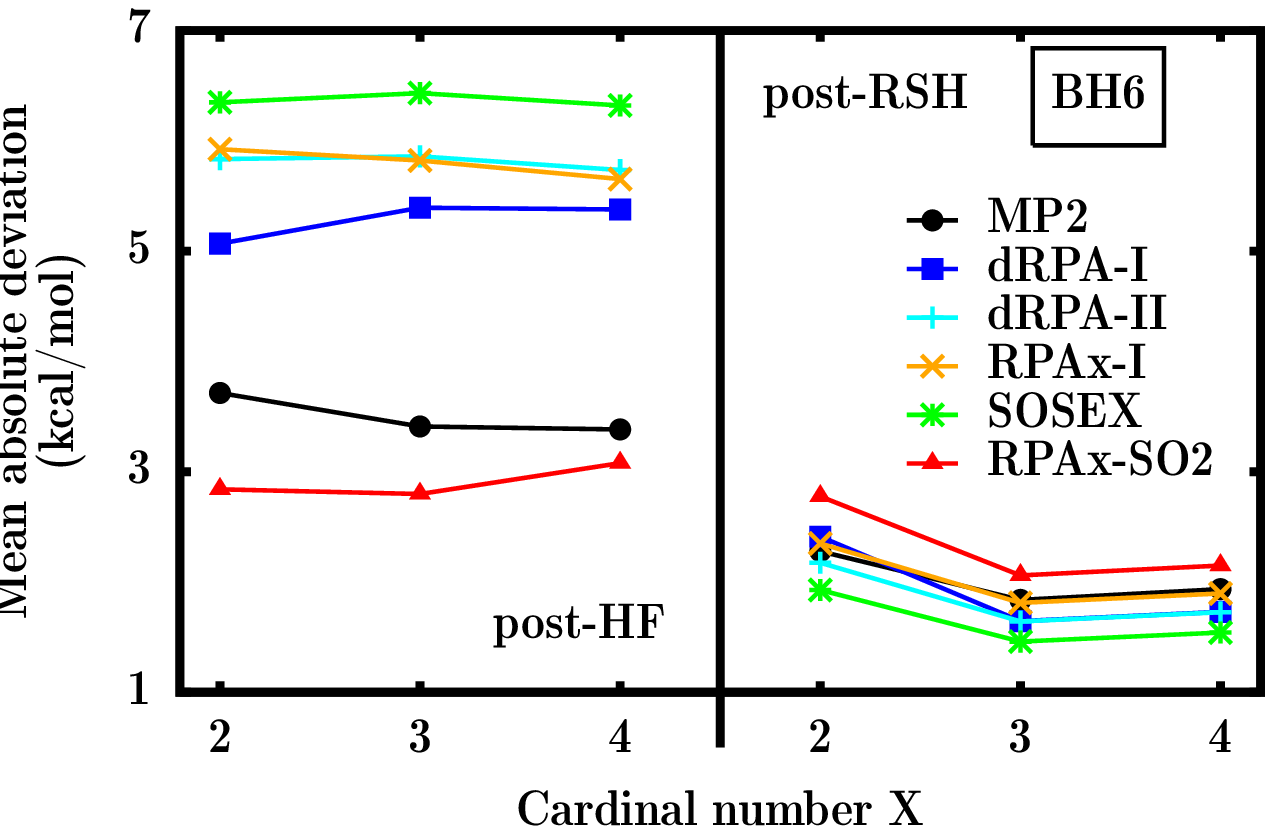}{2cm}{0cm}{0cm}{0cm}{8cm}{\linewidth}
   \label{fig:depBASISBH6}
  \end{minipage}

  \caption{Mean absolute deviations for the AE6 and BH6 datasets as functions of the cardinal number X of the Dunning cc-pVXZ basis sets for full-range post-HF and range-separated post-RSH (srPBE functional and $\mu=0.5$) calculations with different correlation methods (MP2, dRPA-I, dRPA-II, RPAx-I, SOSEX, and RPAx-SO2).
  The reference values are the non-relativistic FC-CCSD(T)/cc-pVQZ-F12 values of Refs.~\onlinecite{Haunschild:12a,Haunschild:13}.
  }
  \label{fig:depBASIS}
\end{minipage}
\end{figure*}
}

\newcommand\tabMUE{
\begin{table}[!htb]
\footnotesize
\caption{\label{tab:MAD} 
Mean absolute deviations (in kcal/mol) for the AE49 and DBH24/08 datasets for range-separated calculations using the srPBE functional and several values of the range-separated parameter ($\mu=0.3$, $\mu=0.5$ and $\mu=0.7$) with different post-RSH long-range correlation methods (MP2, dRPA-I, and RPAx-SO2). The case $\mu=\infty$ corresponds to full-range post-HF calculations. The basis sets are cc-pVQZ for the AE49 dataset and aug-cc-pVQZ for the DBH24/08 dataset. The reference values are the non-relativistic FC-CCSD(T)/cc-pVQZ-F12 values of Ref.~\onlinecite{Haunschild:12b} for the AE49 dataset and the values of Ref.~\onlinecite{Zheng:09} for the DBH24/08 dataset.
}
 \begin{tabularx}{0.48\textwidth}{lRRRRcRRRR}
  \hline\hline
		& \multicolumn{4}{c}{AE49}		&& \multicolumn{4}{c}{DBH24/08}	\\
		  \cline{2-5}				   \cline{7-10}
  $\mu=$	& 0.3  & 0.5  & 0.7    & $\infty$	&& 0.3  & 0.5  & 0.7  & $\infty$\\
  \hline
  MP2		& 6.34 & 5.09 &  7.00  &   5.63		&& 8.28 & 2.94 & 3.45 & 6.17    \\
  dRPA-I	& 6.38 & 6.80 & 12.09  &  23.23		&& 8.29 & 3.01 & 3.89 & 6.95    \\
  RPAx-SO2	& 6.50 & 4.06 &  5.08  &  12.20		&& 5.12 & 2.83 & 3.27 & 6.40    \\
  \hline\hline
 \end{tabularx}
\normalsize
\end{table}
}

\newcommand\figNDist{
\begin{figure*}[!htb]
\centering
\begin{minipage}[t]{.48\linewidth}
   AE49\phantom{/}
   \vskip3mm
  \begin{minipage}[t]{.5\linewidth}
   \fig{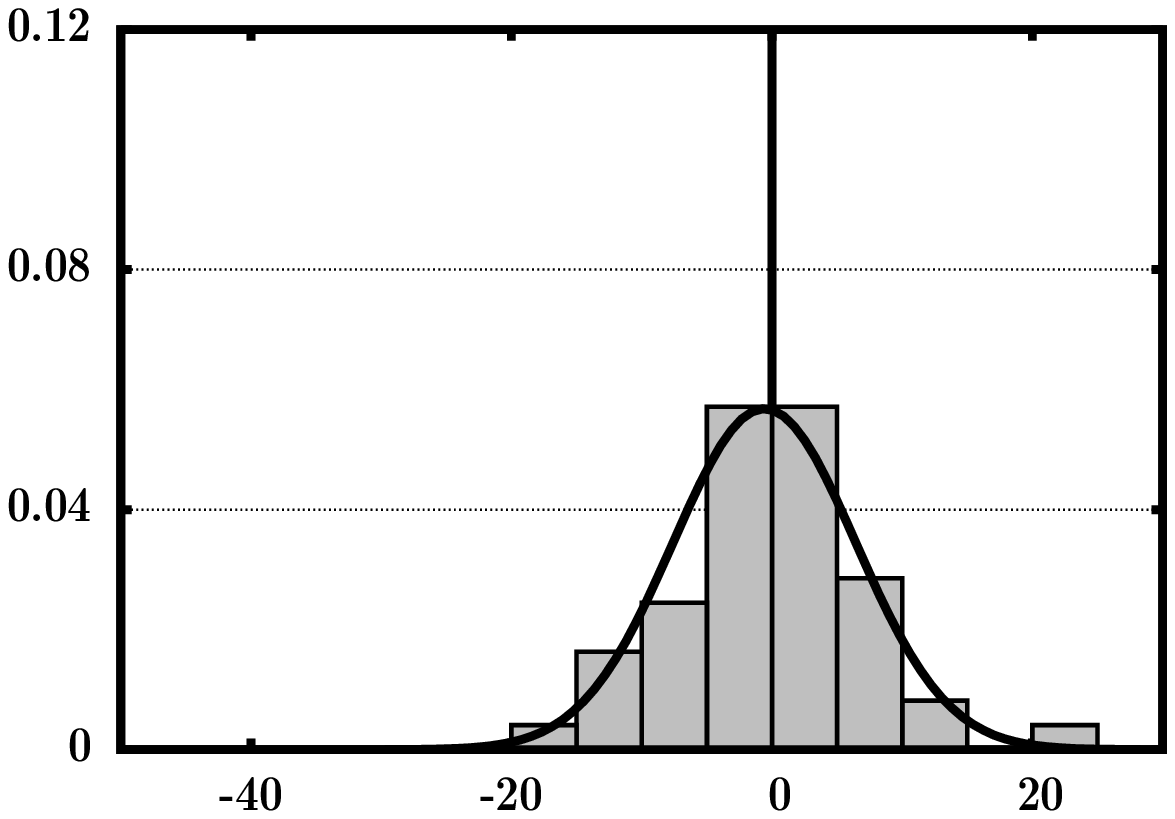}{1cm}{0cm}{0cm}{0cm}{8cm}{\linewidth}
   (a) HF+MP2
   \label{fig:HFMP2AE55}
  \end{minipage}%
  \begin{minipage}[t]{.5\linewidth}
   \fig{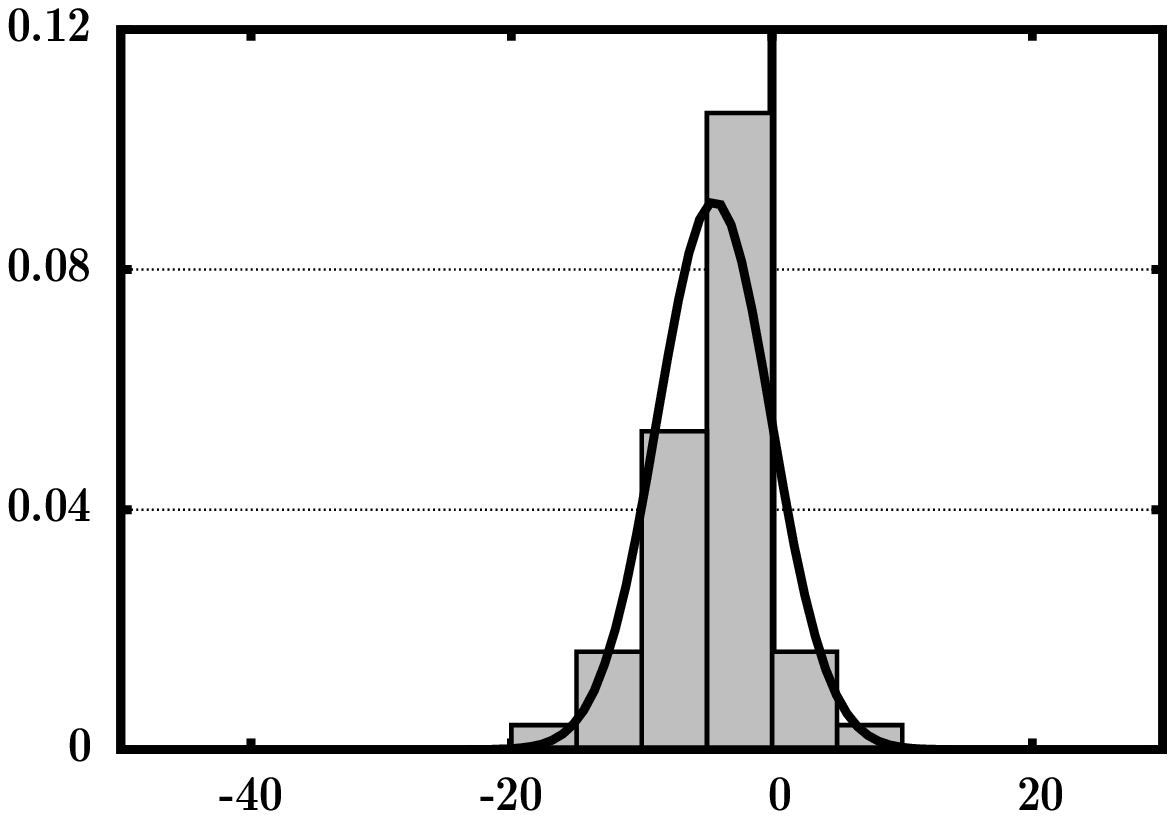}{1cm}{0cm}{0cm}{0cm}{8cm}{\linewidth}
   (b) RSH+MP2
   \label{fig:RSHMP2AE55}
  \end{minipage}
  
  \begin{minipage}[t]{.5\linewidth}
   \fig{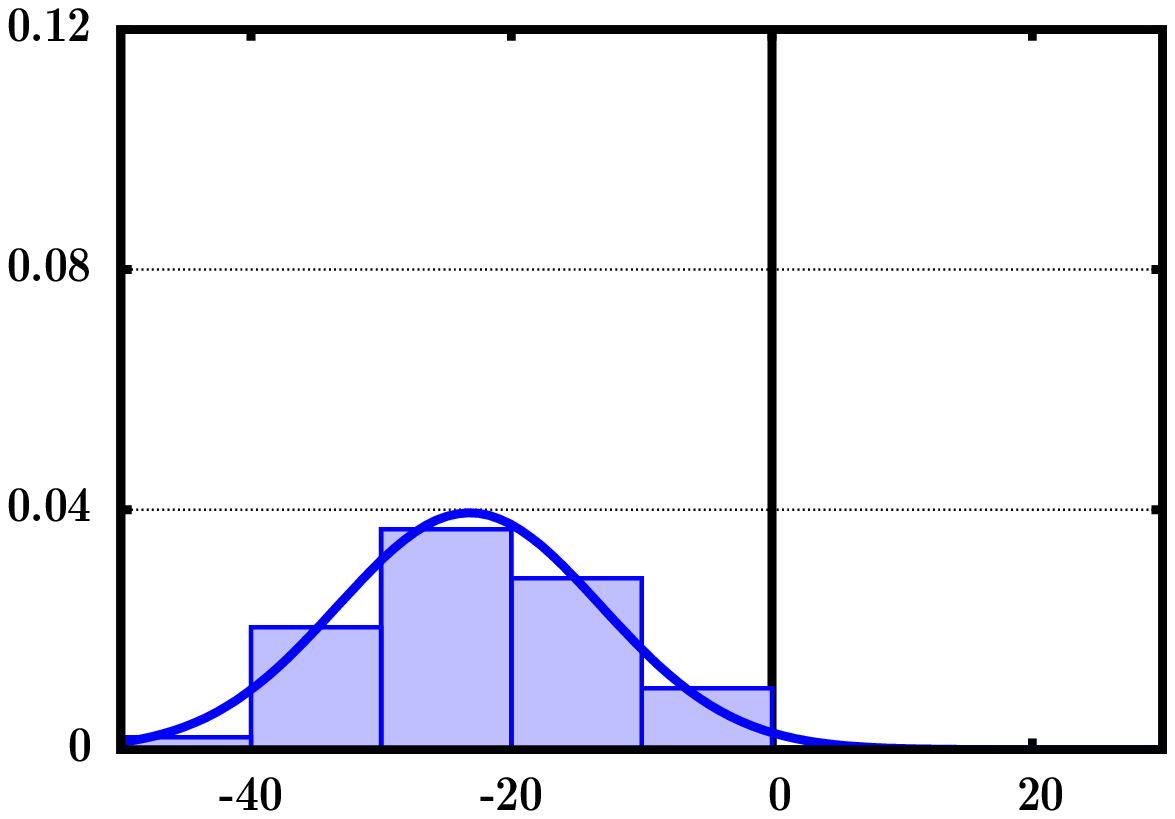}{1cm}{0cm}{0cm}{0cm}{8cm}{\linewidth}
   (c) HF+dRPA-I
   \label{fig:HFdIAE55}
  \end{minipage}%
  \begin{minipage}[t]{.5\linewidth}
   \fig{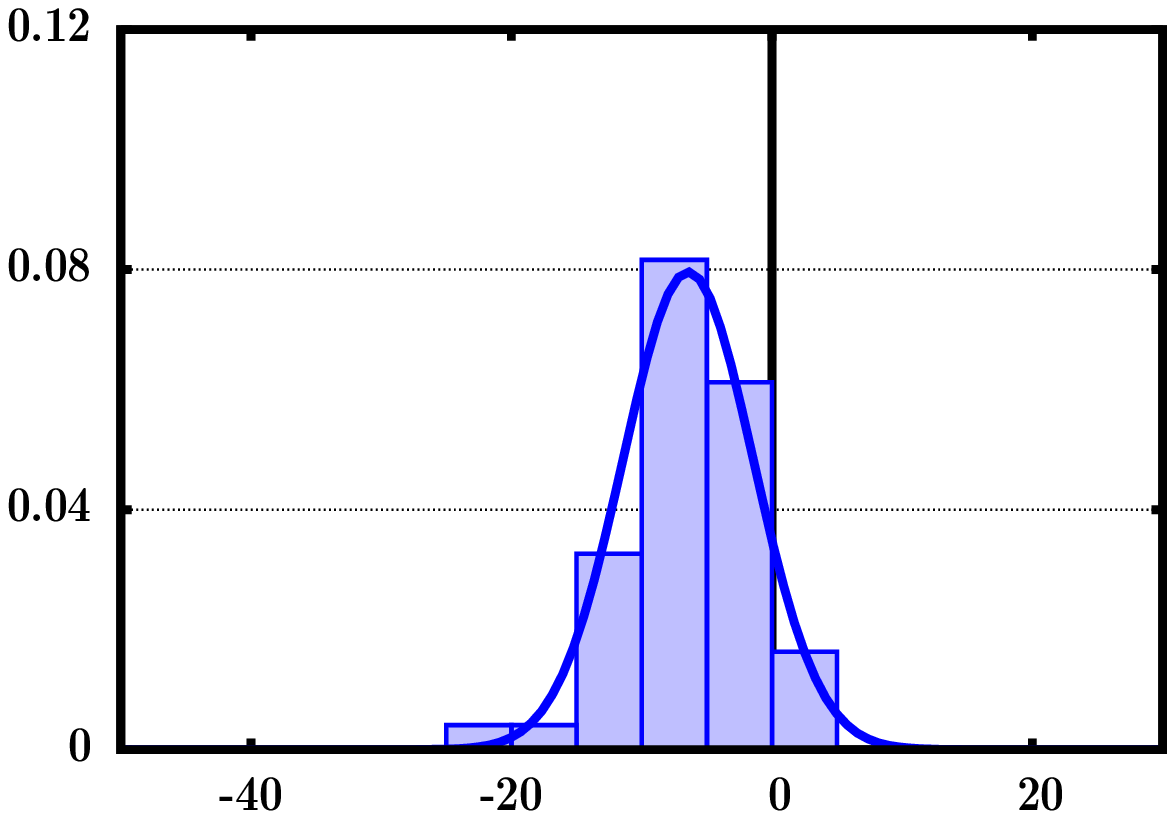}{1cm}{0cm}{0cm}{0cm}{8cm}{\linewidth}
   (d) RSH+dRPA-I
   \label{fig:RSHdIAE55}
  \end{minipage}
  
  \begin{minipage}[t]{.5\linewidth}
   \fig{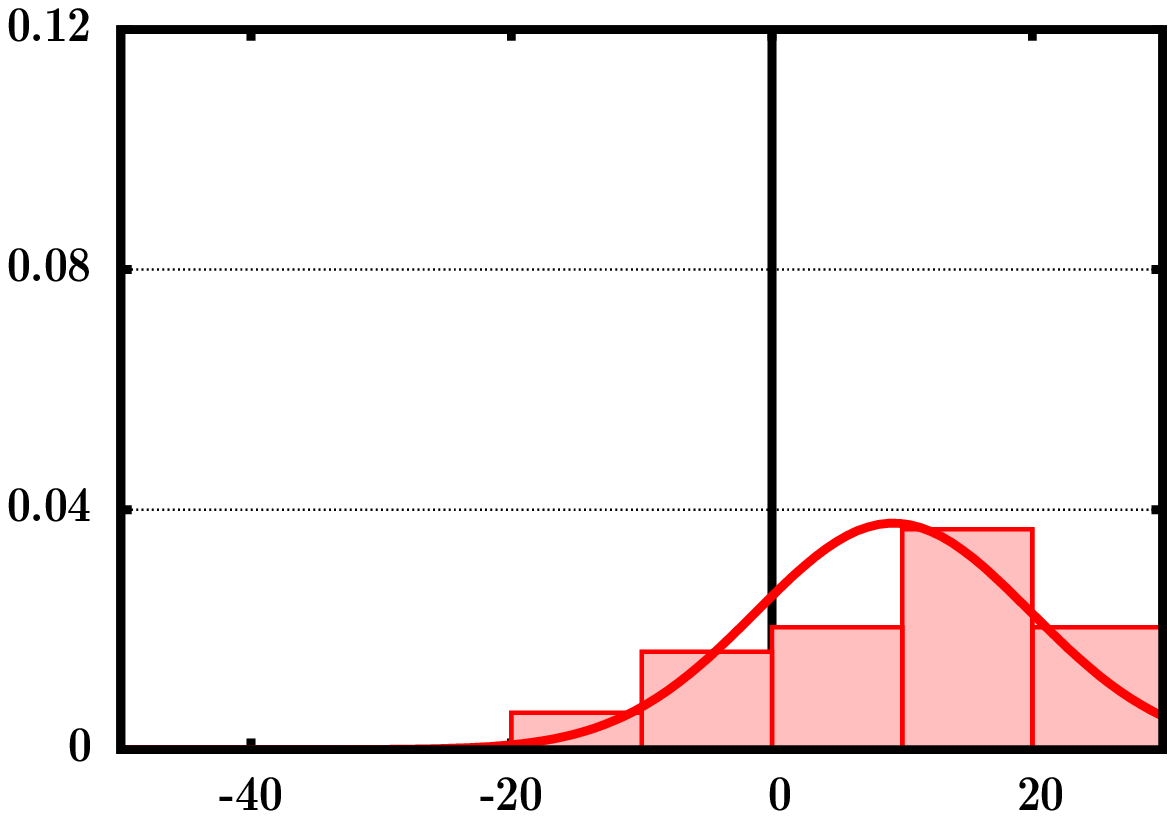}{1cm}{0cm}{0cm}{0cm}{8cm}{\linewidth}
   \label{fig:HFSO2AE55}
   (e) HF+RPAx-SO2
  \end{minipage}%
  \begin{minipage}[t]{.5\linewidth}
   \fig{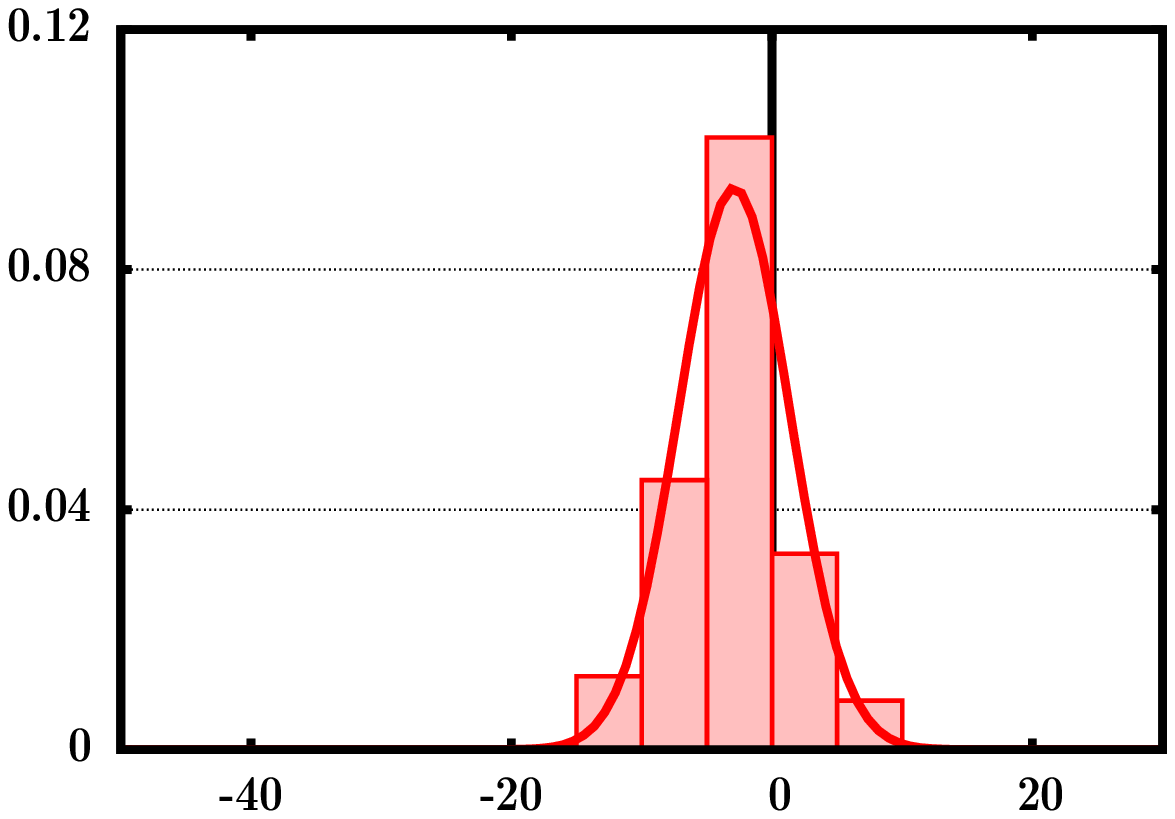}{1cm}{0cm}{0cm}{0cm}{8cm}{\linewidth}
   (f) RSH+RPAx-SO2
   \label{fig:RSHSO2AE55}
  \end{minipage}
  
  \caption{Distribution of the errors (in kcal/mol) obtained with full-range (post-HF) and range-separated (post-RSH) calculations on the AE49 dataset using the cc-pVQZ basis set and $\mu=0.5$. The bins are the distributions of the actual errors and the curves are fitted Gaussian distributions. The reference values are the non-relativistic FC-CCSD(T)/cc-pVQZ-F12 values of Ref.~\onlinecite{Haunschild:12b}.
  }
  \label{fig:NDistAE55}
\end{minipage}\qquad%
\begin{minipage}[t]{.48\linewidth}
   DBH24/08
   \vskip3mm
  \begin{minipage}[t]{.5\linewidth}
   \fig{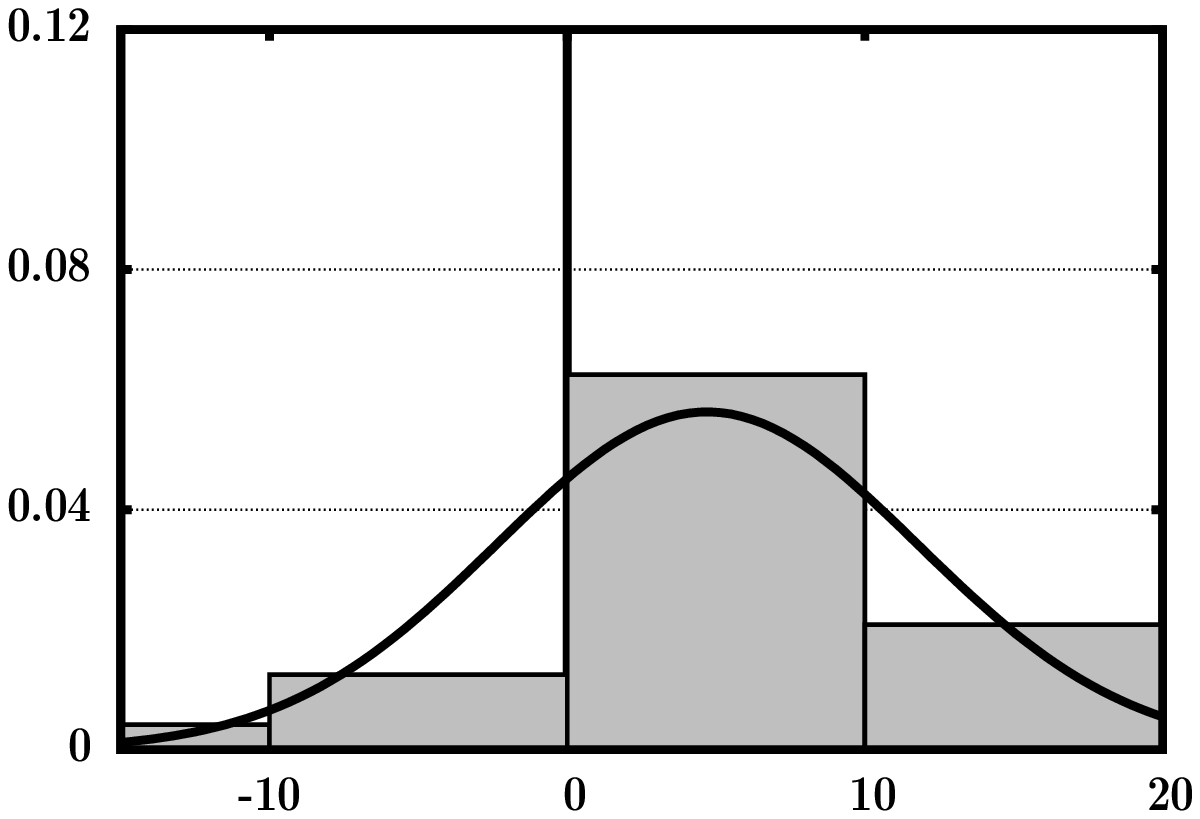}{1cm}{0cm}{0cm}{0cm}{8cm}{\linewidth}
   (a) HF+MP2
   \label{fig:HFMP2BH24}
  \end{minipage}%
  \begin{minipage}[t]{.5\linewidth}
   \fig{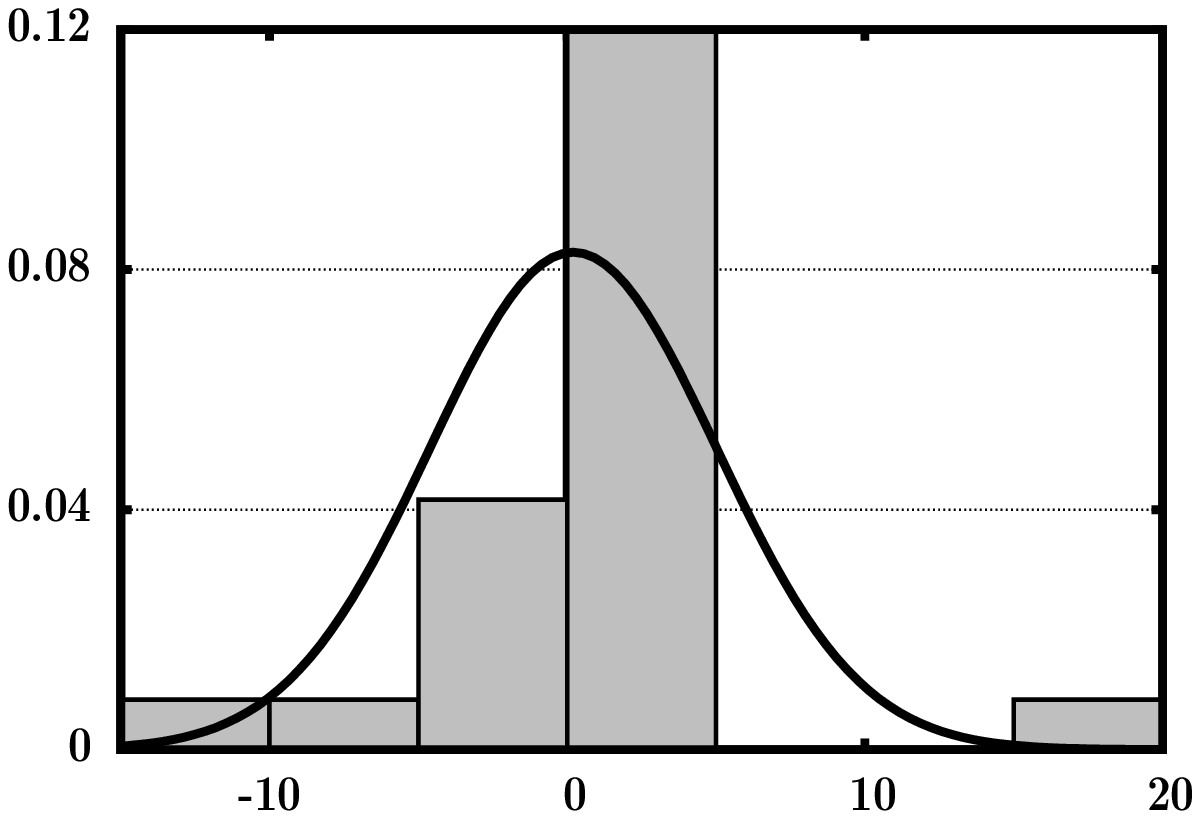}{1cm}{0cm}{0cm}{0cm}{8cm}{\linewidth}
   (b) RSH+MP2
   \label{fig:RSHMP2BH24}
  \end{minipage}
  
  \begin{minipage}[t]{.5\linewidth}
   \fig{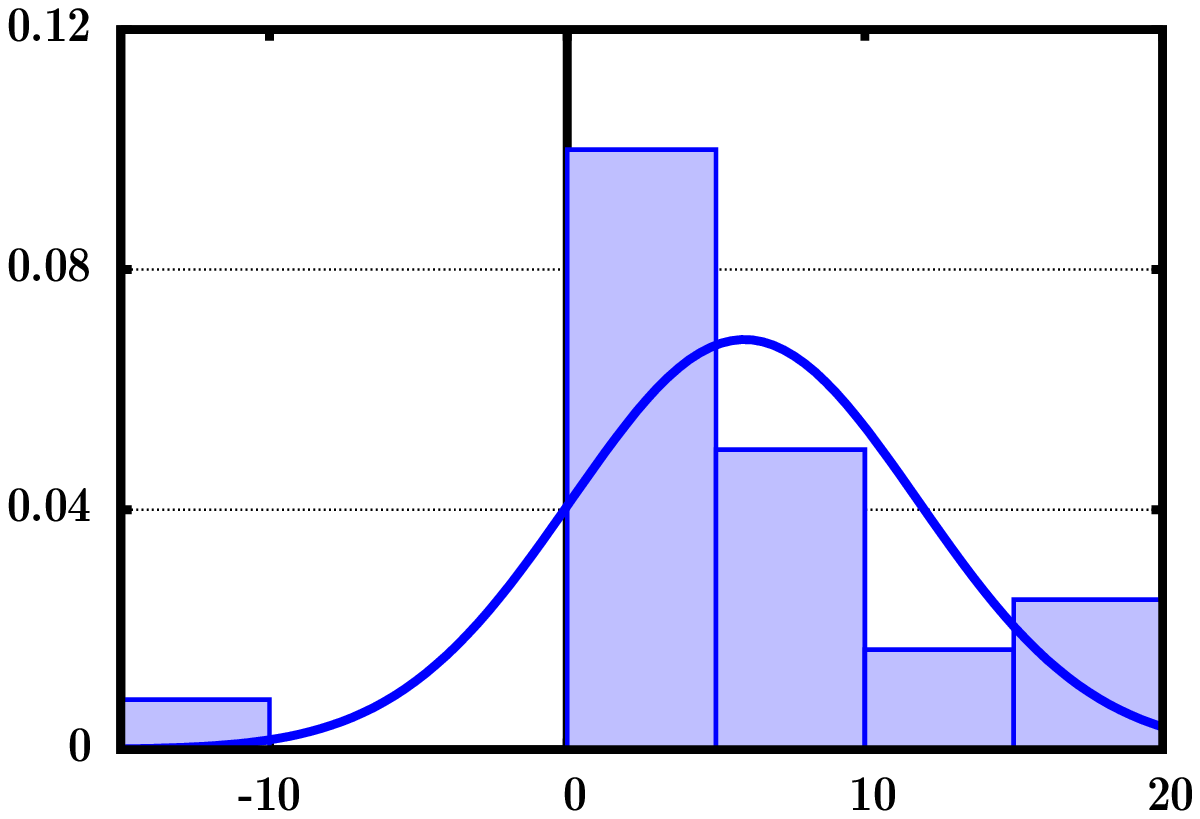}{1cm}{0cm}{0cm}{0cm}{8cm}{\linewidth}
   (c) HF+dRPA-I
   \label{fig:HFdIBH24}
  \end{minipage}%
  \begin{minipage}[t]{.5\linewidth}
   \fig{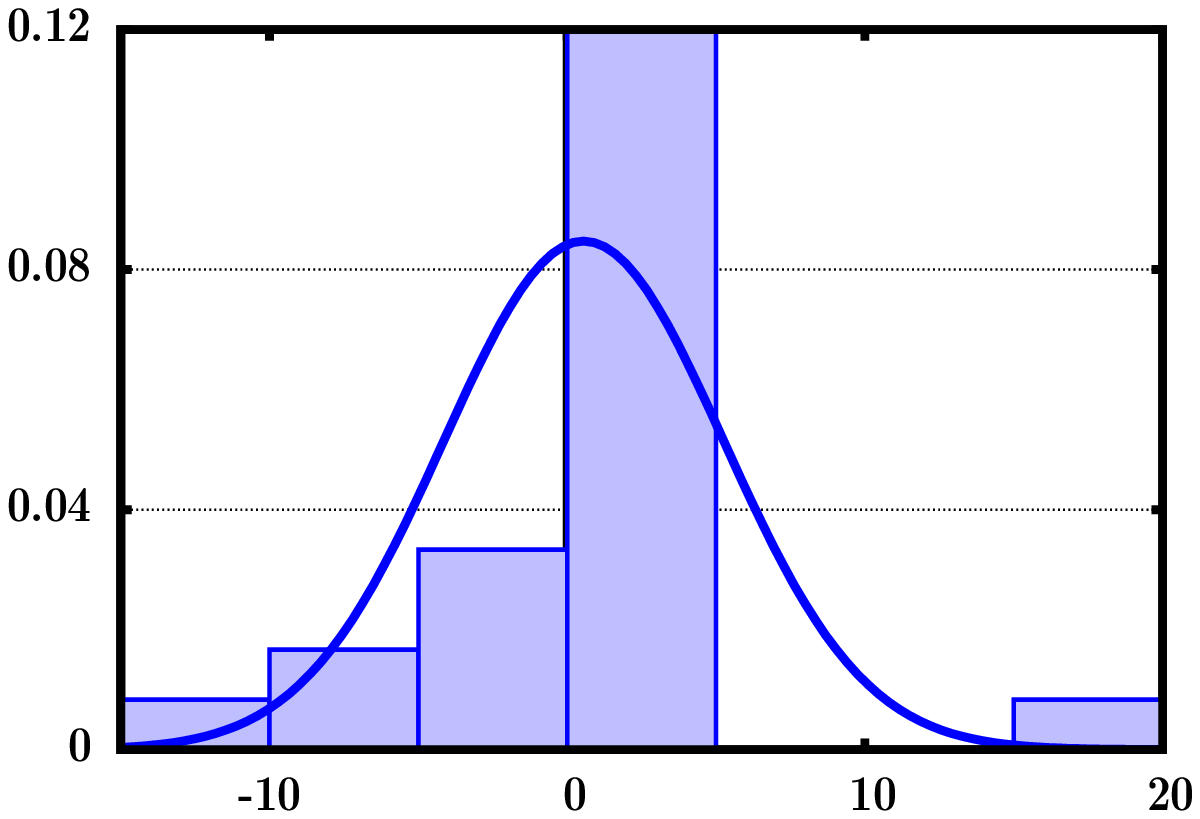}{1cm}{0cm}{0cm}{0cm}{8cm}{\linewidth}
   (d) RSH+dRPA-I
   \label{fig:RSHdIBH24}
  \end{minipage}
  
  \begin{minipage}[t]{.5\linewidth}
   \fig{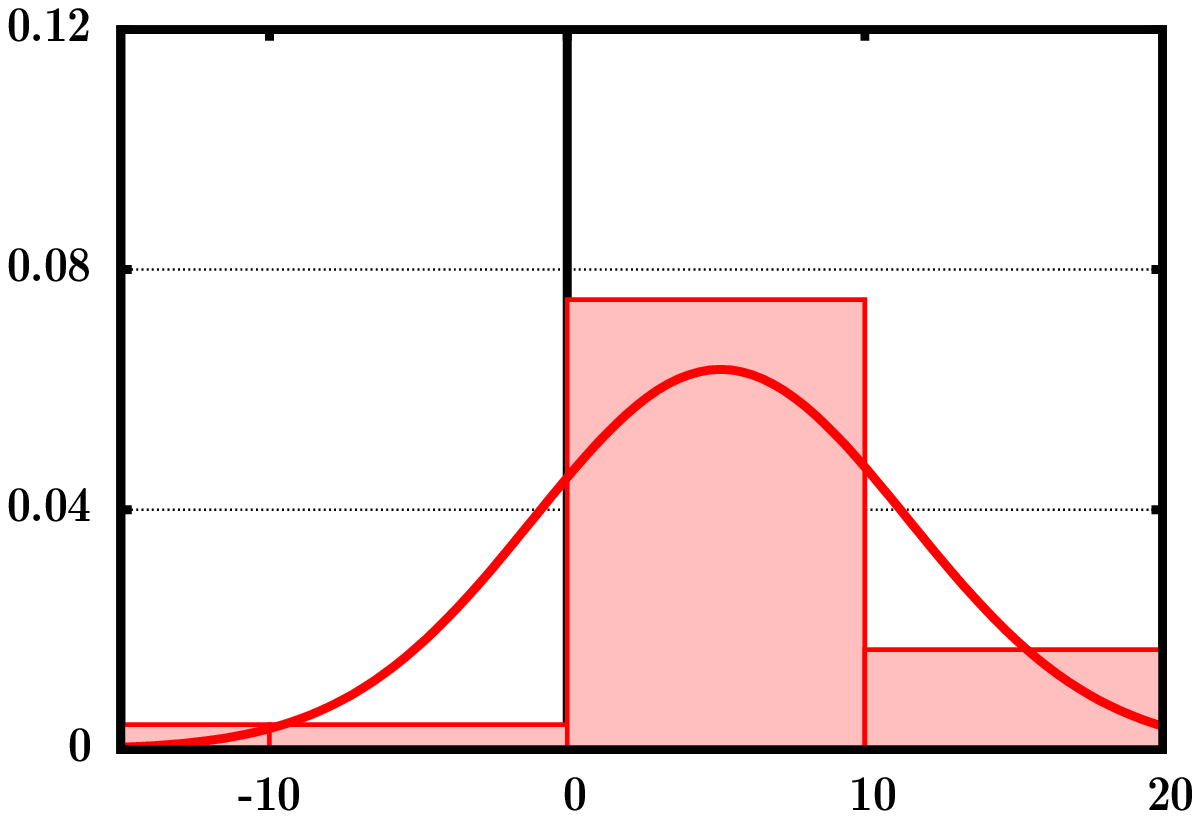}{1cm}{0cm}{0cm}{0cm}{8cm}{\linewidth}
    (e) HF+RPAx-SO2
   \label{fig:HFSO2BH24}
  \end{minipage}%
  \begin{minipage}[t]{.5\linewidth}
   \fig{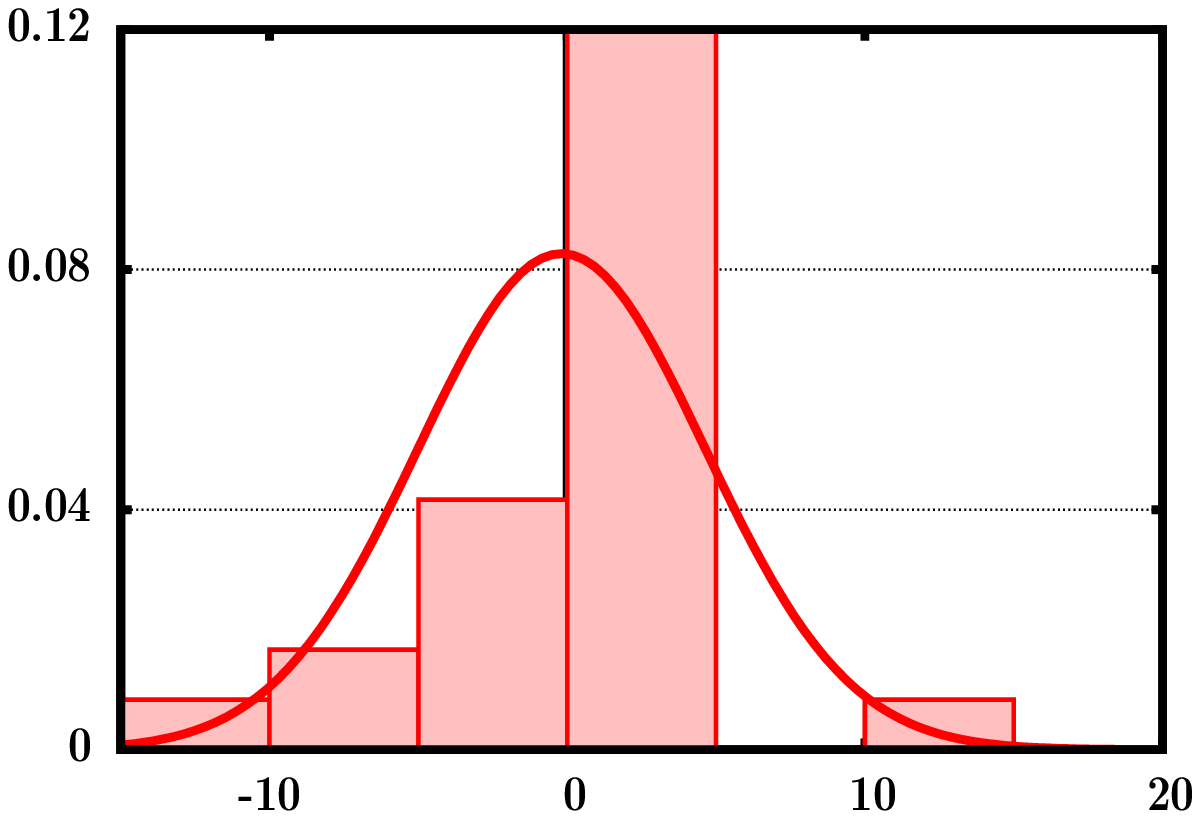}{1cm}{0cm}{0cm}{0cm}{8cm}{\linewidth}
   (f) RSH+RPAx-SO2
   \label{fig:RSHSO2BH24}
  \end{minipage}
  \caption{Distribution of the errors (in kcal/mol) obtained with full-range (post-HF) and range-separated (post-RSH) calculations on the DBH24/08 dataset using the aug-cc-pVQZ basis set and $\mu=0.5$. The bins are the distributions of the actual errors and the curves are fitted Gaussian distributions. The reference values are taken from Ref.~\onlinecite{Zheng:09}.}
  \label{fig:NDistBH24}
\end{minipage}
\end{figure*}
}

\newcommand\figMeanVar{
\begin{figure*}[!htb]
\centering
\begin{minipage}[t]{.5\linewidth}
 \fig{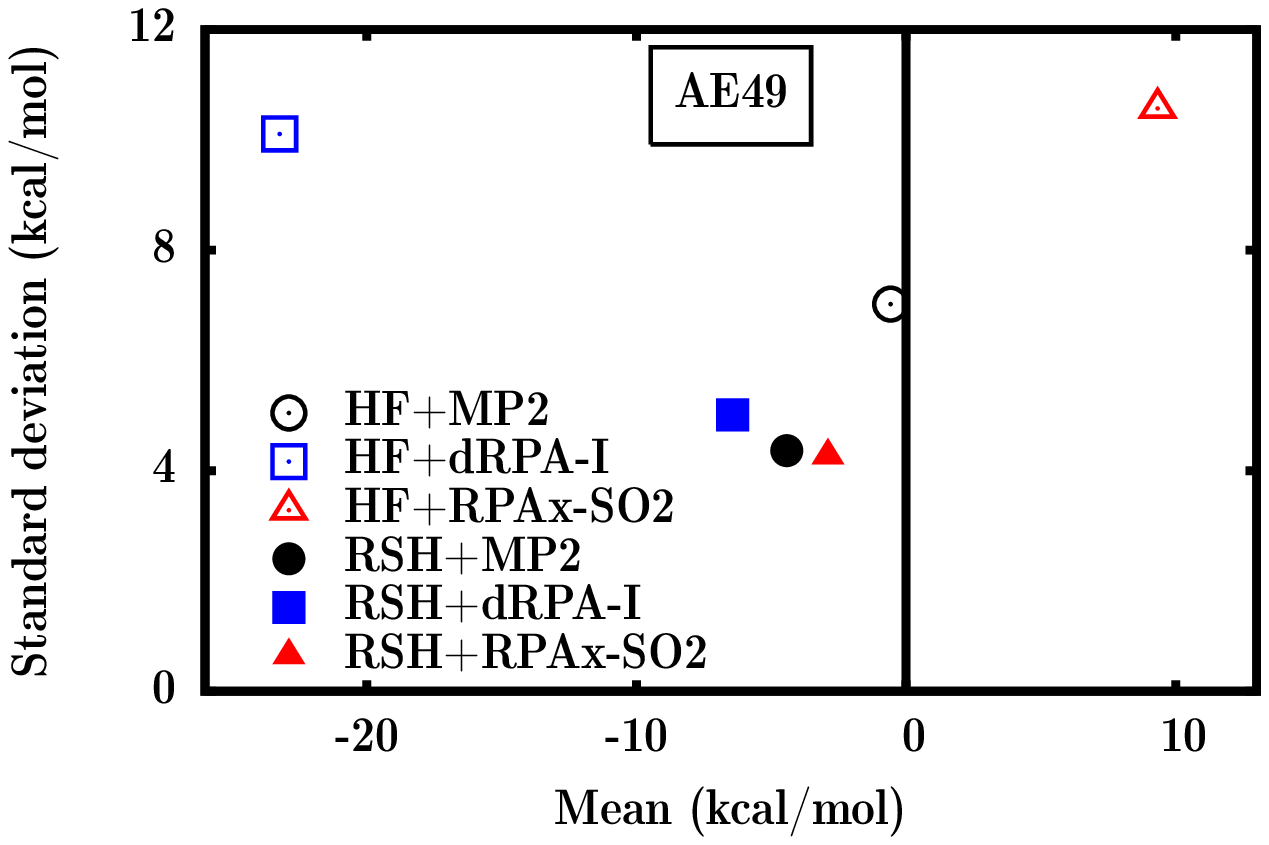}{25mm}{0cm}{0cm}{0cm}{8cm}{\linewidth}
 \label{fig:MeanVarAE55}
\end{minipage}%
\begin{minipage}[t]{.5\linewidth}
 \fig{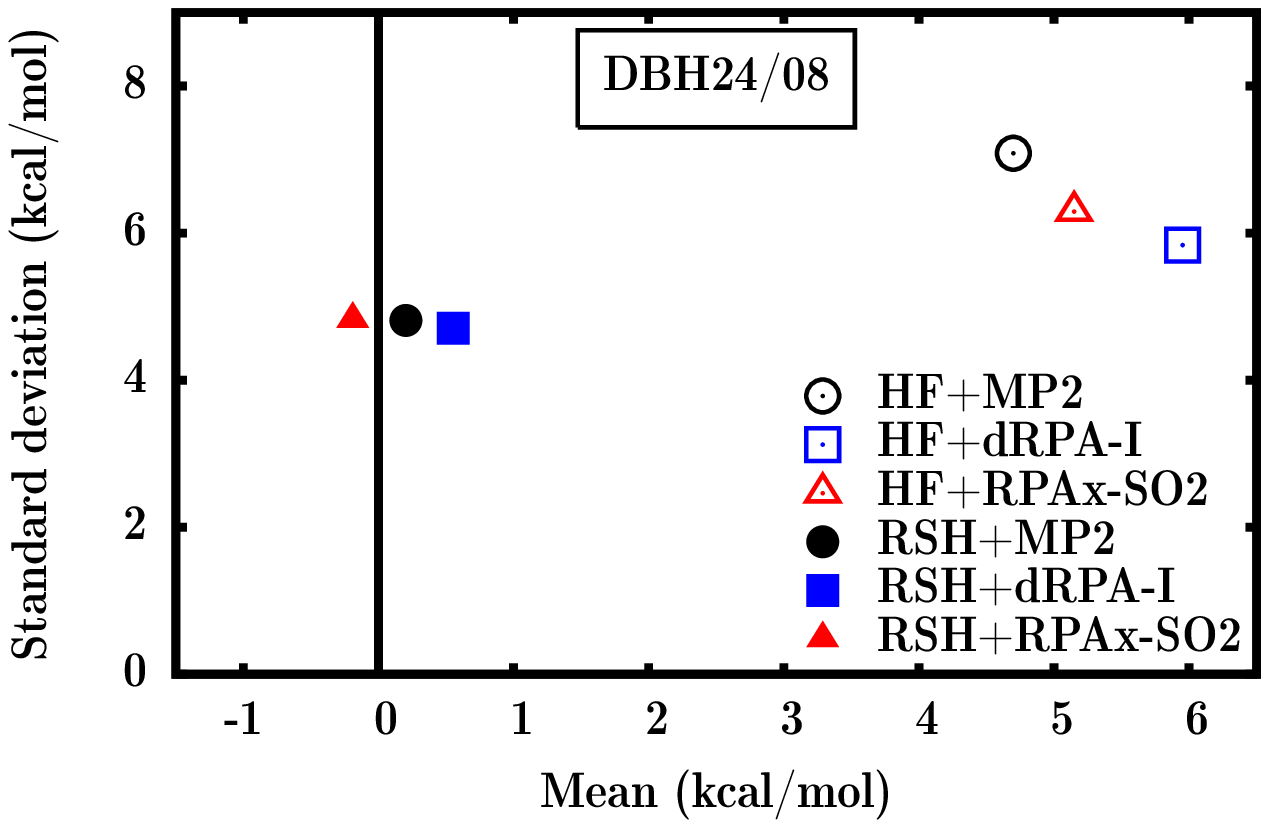}{25mm}{0cm}{0cm}{0cm}{8cm}{\linewidth}
 \label{fig:MeanVarBH24}
\end{minipage}

\caption{\label{fig:MeanVar}
Representation of the means $m$ and standard deviations $\sigma$ (in kcal/mol)
of the Gaussian distributions fitted to the distribution of errors seen in
Figures~\FloatRef{fig:NDistAE55} and~\FloatRef{fig:NDistBH24} for full-range (post-HF) and range-separated (post-RSH) calculations
on the AE49 and DBH24/08 datasets.
}
\end{figure*}
}

\begin{document}

\title{Spin-unrestricted random-phase approximation with range separation: Benchmark on atomization energies and reaction barrier heights}

\author{Bastien Mussard}\email{bastien.mussard@upmc.fr}
\affiliation{Sorbonne Universit\'es, UPMC Univ Paris 06, Institut du Calcul et de la Simulation, F-75005, Paris, France}
\affiliation{Sorbonne Universit\'es, UPMC Univ Paris 06, UMR 7616, Laboratoire de Chimie Th\'eorique, F-75005 Paris, France}
\affiliation{CNRS, UMR 7616, Laboratoire de Chimie Th\'eorique, F-75005 Paris, France}
\author{Peter Reinhardt}
\affiliation{Sorbonne Universit\'es, UPMC Univ Paris 06, UMR 7616, Laboratoire de Chimie Th\'eorique, F-75005 Paris, France}
\affiliation{CNRS, UMR 7616, Laboratoire de Chimie Th\'eorique, F-75005 Paris, France}
\author{J\'anos G. \'Angy\'an}
\affiliation{CRM2, Institut Jean Barriol, Universit\'e de Lorraine, F-54506 Vand{\oe}uvre-l\'es-Nancy, France}
\affiliation{CRM2, Institut Jean Barriol, CNRS, F-54506 Vand{\oe}vre-l\'es-Nancy, France}
\author{Julien Toulouse}\email{julien.toulouse@upmc.fr}
\affiliation{Sorbonne Universit\'es, UPMC Univ Paris 06, UMR 7616, Laboratoire de Chimie Th\'eorique, F-75005 Paris, France}
\affiliation{CNRS, UMR 7616, Laboratoire de Chimie Th\'eorique, F-75005 Paris, France}

\begin{abstract}
We consider several spin-unrestricted random-phase approximation (RPA) variants for calculating correlation energies, with and without range separation, and test them on datasets of atomization energies and reaction barrier heights. We show that range separation greatly improves the accuracy of all RPA variants for these properties. Moreover, we show that a RPA variant with exchange, hereafter referred to as RPAx-SO2, first proposed by Szabo and Ostlund [A. Szabo and N. S. Ostlund, J. Chem. Phys. {\bf 67}, 4351 (1977)] in a spin-restricted closed-shell formalism, and extended here to a spin-unrestricted formalism, provides on average the most accurate range-separated RPA variant for atomization energies and reaction barrier heights. Since this range-separated RPAx-SO2 method had already been shown to be among the most accurate range-separated RPA variants for weak intermolecular interactions [J. Toulouse, W. Zhu, A. Savin, G. Jansen, and J. G. \'Angy\'an, J. Chem. Phys. {\bf 135}, 084119 (2011)], this works confirms range-separated RPAx-SO2 as a promising method for general chemical applications.
\end{abstract}

\maketitle

\section{Introduction}
\label{sec:intro}

There has recently been a revived interest in the random-phase approximation (RPA) for the calculation of electron correlation energies of atomic, molecular and solid-state systems (see Refs.~\onlinecite{EshBatFur-TCA-12,RenRinJoaSch-JMS-12} for recent reviews). One of the advantages of RPA is its good description of dispersion interactions at large separation~\cite{Dobson:01,Dobson:05,DobGou-JPCM-12}, whereas two important disadvantages are its poor description of short-range electron correlations~\cite{Yan:00} and its slow Gaussian basis convergence~\cite{Furche:01b,EshFur-JCP-12,FabDel-TCA-12}. These two limitations can be overcome by the range-separation approach (see, e.g., Ref.~\onlinecite{TouColSav-PRA-04}) which allows one to combine a short-range density-functional approximation with a long-range RPA-type approximation~\cite{TouGerJanSavAng-PRL-09,JanHenScu-JCP-09,JanHenScu-JCP-09b,ZhuTouSavAng-JCP-10,TouZhuAngSav-PRA-10,TouZhuSavJanAng-JCP-11,AngLiuTouJan-JCTC-11,IreHenScu-JCP-11,CheMusAngRei-CPL-12,MusSzaAng-JCTC-14}.

As a result of the increasing interest in RPA, there are now several RPA formulations in which RPA equations can be derived, namely the adiabatic connection~\cite{Furche:01b,Fuchs:02,Angyan:11}, dielectric matrix~\cite{Nozieres:58a,Harl:08,Nguyen:09}, plasmon~\cite{McLachlan:64,Scuseria:08} formula and ring coupled-cluster doubles~\cite{SzaOst-JCP-77,Scuseria:08,TouZhuSavJanAng-JCP-11} formulations. Moreover, within these formulations, many variants of RPA (\textit{e.g.}, direct RPA~\cite{Scuseria:08}, RPA with exact Hartree-Fock (HF) exchange~\cite{SzaOst-JCP-77,Oddershede:78,BatFur-JCP-13}, RPA with exact Kohn-Sham exchange~\cite{Hesselmann:10}) can be defined (see also Refs.~\onlinecite{AngLiuTouJan-JCTC-11,JanLiuAng-JCP-10,TouZhuSavJanAng-JCP-11,Hes-TCC-14}). 

Recently, in a series of papers, we have studied the performance of these RPA variants for closed-shell systems, especially in the range-separated framework.
Within the ring coupled-cluster doubles formulation, it was found that a spin-restricted closed-shell RPA variant with exchange, first proposed by Szabo and Ostlund
(see equation (3.22) of Ref.~\onlinecite{SzaOst-JCP-77})
and referred to as Szabo-Ostlund's second formula (RPAx-SO2) in Ref.~\onlinecite{TouZhuSavJanAng-JCP-11}, was one of the best RPA variants for correlation energies, chemical reaction energies, and weak intermolecular binding energies for closed-shell systems with or without range separation~\cite{Hes-JCP-11,TouZhuSavJanAng-JCP-11}. In view of future applications, like the simulation of reactivity in condensed phases, where different kinds of intermolecular and intramolecular forces are acting simultaneously in a concomitant manner, it is interesting to know whether we can expect a similar degree of reliability in the description of chemical transformations as for weak non-covalent interactions.  

The purpose of this work is thus to further test the performance of the RPAx-SO2 variant on more general thermochemistry properties, namely atomization energies and reaction barrier heights, now involving open-shell systems. To this end, we have extended and implemented the RPAx-SO2 variant, as well as other RPA variants, in a spin-unrestricted formalism with and without range-separation. Another important domain of application of open-shell range-separated RPA may consist of the study of transition metal compounds, which is out of the scope of the present study and will be the subject of forthcoming publications.

The paper is organized as follows. In Section~\ref{sec:theory}, we offer an overview of the RPA variants in the adiabatic connection and ring coupled-cluster doubles formulations in a spin-unrestricted formalism. After giving computational details in Section~\ref{sec:compdetails}, we present in Section~\ref{sec:results} the results of the different RPA variants with and without range-separation on datasets of atomization energies and reaction barrier heights. A detailed statistical analysis of the results is given. Finally, Section~\ref{sec:conclusion} contains our conclusions.

\section{Spin-unrestricted RPA correlation energy expressions}
\label{sec:theory}

We start by giving the RPA correlation energy expressions used in this work, using real-valued canonical spin-orbitals in a spin-unrestricted formalism for general open-shell systems, breaking spatial and $\hat{S}^2$-spin symmetry (but not $\hat{S}_z$-spin symmetry).

The RPA correlation energy expressions can be derived in different \textit{formulations} (see Ref.~\onlinecite{AngLiuTouJan-JCTC-11}). One way to appreciate the links between the formulations is to consider the adiabatic-connection fluctuation-dissipation theorem expression for the correlation energy~\cite{Langreth:75,Langreth:77}, which involves integrations over both a frequency and a coupling constant. An analytical integration over the frequency variable followed by a numerical integration over the coupling constant yields the \textit{adiabatic-connection formulation}, while an analytic integration over the coupling constant followed by a numerical integration over the frequency yields the \textit{dielectric formulation}. In certain cases, the two integrations can be carried out analytically which leads to two other formulations: the \textit{plasmon-formula formulation} and the \textit{ring coupled-cluster doubles formulation}. Depending on the approximations used, the correlation energies derived in these four formulations are in general not equivalent. Here we consider only the adiabatic-connection and ring coupled-cluster doubles formulations.

In the adiabatic-connection formulation, the RPA correlation energy is written as the integral over the coupling constant $\alpha$ of the trace of the correlation part of a two-particle density matrix $\b{P}_{c,\alpha}$ contracted with a matrix $\b{B}_1$ of two-electron integrals at $\alpha=1$. Depending whether the calculation of $\b{P}_{c,\alpha}$ includes only a \textit{direct} term (dRPA) or also a HF \textit{exchange} term (RPAx), and whether the two-electron integrals in $\b{B}_1$ are \textit{non-antisymmetrized} (I) or \textit{antisymmetrized} (II), four RPA correlation energy expressions can be defined: the dRPA-I variant (also called dRPA or just RPA in the density-functional theory literature)~\cite{Langreth:75,Langreth:77}
\begin{equation}
E_\c^\text{dRPA-I} = \frac{1}{2} \int_0^{1} \d \alpha \;\tr \left[ \b{B}^\I_1  \; \b{P}_{c,\alpha}^\dRPA \right],
\label{eq:ACdRPAI}
\end{equation}
the dRPA-II variant~\cite{AngLiuTouJan-JCTC-11}
\begin{equation}
E_\c^\text{dRPA-II} = \frac{1}{2} \int_0^{1} \d \alpha \;\tr \left[ \b{B}^\II_1 \; \b{P}_{c,\alpha}^\dRPA \right],
\label{eq:ACdRPAII}
\end{equation}
the RPAx-I variant~\cite{TouGerJanSavAng-PRL-09,TouZhuAngSav-PRA-10}
\begin{equation}
E_\c^\text{RPAx-I} = \frac{1}{2} \int_0^{1} \d \alpha \;\tr \left[ \b{B}^\I_1  \; \b{P}_{c,\alpha}^\RPAx \right],
\label{eq:ACRPAxI}
\end{equation}
and the RPAx-II variant~\cite{McLachlan:64,AngLiuTouJan-JCTC-11}
\begin{equation}
E_\c^\text{RPAx-II} = \frac{1}{4} \int_0^{1} \d \alpha \;\tr \left[ \b{B}^\II_1  \; \b{P}_{c,\alpha}^\RPAx \right].
\label{eq:ACRPAxII}
\end{equation}
In all these cases, $\b{P}_{c,\alpha}$ is calculated as~\cite{Furche:01b}
\begin{equation}
\b{P}_{c,\alpha} = (\b{A}_\alpha-\b{B}_\alpha)^{1/2} \; \b{M}_\alpha^{-1/2} \; (\b{A}_\alpha-\b{B}_\alpha)^{1/2} -\b{1},
\end{equation}
where $\b{M}_\alpha = (\b{A}_\alpha-\b{B}_\alpha)^{1/2} \; (\b{A}_\alpha+\b{B}_\alpha) \; (\b{A}_\alpha-\b{B}_\alpha)^{1/2}$. The dRPA density matrix $\b{P}_{c,\alpha}^\dRPA$ is obtained using the following definition for the matrices $\b{A}_\alpha$ and $\b{B}_\alpha$, for an arbitrary coupling constant $\alpha$,
\begin{equation}
\left(A_\alpha^\I \right)_{ia,jb} = (\varepsilon_a - \varepsilon_i) \delta_{ij} \delta_{ab} + \alpha \; \braket{ib}{aj},
\end{equation}
\begin{equation}
\left(B_\alpha^\I \right)_{ia,jb} = \alpha \; \braket{ab}{ij},
\end{equation}
where $i,j$ and $a,b$ refer to occupied and virtual spin-orbitals, respectively, $\varepsilon_i$ and $\varepsilon_a$ are the spin-orbital energies, and $\braket{ib}{aj}$ and $\braket{ab}{ij}$ are non-antisymmetrized two-electron integrals. Similarly, the RPAx density matrix $\b{P}_{c,\alpha}^\RPAx$ is obtained using the matrices
\begin{equation}
\left(A_\alpha^\II \right)_{ia,jb} = (\varepsilon_a - \varepsilon_i) \delta_{ij} \delta_{ab} + \alpha \; \bra{ib}\ket{aj},
\end{equation}
\begin{equation}
\left(B_\alpha^\II \right)_{ia,jb} = \alpha \; \bra{ab}\ket{ij},
\end{equation}
where $\bra{ib}\ket{aj}=\braket{ib}{aj}-\braket{ib}{ja}$ and $\bra{ab}\ket{ij}=\braket{ab}{ij}-\braket{ab}{ji}$ are antisymmetrized two-electron integrals.

In the ring coupled-cluster doubles formulation, the key quantity replacing $\b{P}_{c,\alpha}$ is the matrix of double-excitation amplitudes $\b{T}$ given by the Riccati equation (at $\alpha=1$)~\cite{Scuseria:08}
\begin{equation}
\b{B}_1 + \b{A}_1 \; \b{T} + \b{T} \; \b{A}_1 +  \b{T} \; \b{B}_1 \; \b{T} = \b{0}.
\label{Riccati}
\end{equation}
The dRPA-I and RPAx-II correlation energies are then obtained by contracting the amplitudes with two-electron integrals~\cite{Scuseria:08}
\begin{equation}
E_c^\text{dRPA-I} = \frac{1}{2} \tr \left[ \b{B}^\I_1  \; \b{T}^\dRPA \right],
\label{EcdRPAI-rCCD}
\end{equation}
\begin{equation}
E_c^\text{RPAx-II} = \frac{1}{4} \tr \left[ \b{B}^\II_1  \; \b{T}^\RPAx \right],
\label{EcRPAxII-rCCD}
\end{equation}
where $\b{T}^\dRPA$ is obtained using the matrices $\b{A}_1^\I$ and $\b{B}_1^\I$ in Eq.~(\ref{Riccati}), and $\b{T}^\RPAx$ is obtained using the matrices $\b{A}_1^\II$ and $\b{B}_1^\II$ in Eq.~(\ref{Riccati}). We emphasize that the correlation energies given by Eqs.~(\ref{EcdRPAI-rCCD}) and~(\ref{EcRPAxII-rCCD}) are identical to the ones given by Eqs.~(\ref{eq:ACdRPAI}) and~(\ref{eq:ACRPAxII}). Additionally, two other RPA correlation energy variants can be defined in this formulation: a dRPA variant with second-order screened exchange (SOSEX)~\cite{GruMarHarSchKre-JCP-09,PaiJanHenScuGruKre-JCP-10,TouZhuSavJanAng-JCP-11}
\begin{equation}
E_c^\text{SOSEX} = \frac{1}{2} \tr \left[ \b{B}^\II_1  \; \b{T}^\dRPA \right],
\label{eq:rCCDSOSEX}
\end{equation}
and a RPAx variant corresponding to Szabo-Ostlund's second formula (SO2)
(see Ref.~\onlinecite{TouZhuSavJanAng-JCP-11} and equation (3.22) in Ref.~\onlinecite{SzaOst-JCP-77})
\begin{equation}
E_c^\text{RPAx-SO2} = \frac{1}{2} \tr \left[ \b{B}^\I_1  \; \b{T}^\RPAx \right].
\label{eq:rCCDSO2}
\end{equation}
The expressions of these last two RPA correlation energy variants are similar but not equivalent to the variants in Eqs.~(\ref{eq:ACdRPAII}) and~(\ref{eq:ACRPAxI}). The RPAx-SO2 correlation energy expression in Eq.~(\ref{eq:rCCDSO2}) is an obvious extension of the closed-shell Szabo-Ostlund's expression to a spin-unrestricted formalism, and is original to this work. We note that there is no obvious extension of the closed-shell RPAx variant corresponding to Szabo-Ostlund's first formula (SO1) 
(see Ref.~\onlinecite{TouZhuSavJanAng-JCP-11} and equation (3.20) in Ref.~\onlinecite{SzaOst-JCP-77})
to a spin-unrestricted formalism. At second order in the electron-electron interaction, all the RPA correlation energy variants reduce to the second-order M{\o}ller-Plesset (MP2) correlation energy expression~\cite{AngLiuTouJan-JCTC-11}, except for the dRPA-I variant which reduces to direct MP2~\cite{JanScu-PCCP-09}.

The dRPA-I, dRPA-II, and SOSEX variants are free from instabilities by construction. In a spin-restricted closed-shell formalism, the RPAx-I and RPAx-SO2 variants involve only spin-singlet excitations and are thus not subject to triplet instabilities, while the RPAx-II variant involves both spin-singlet and spin-triplet excitations and are thus prone to triplet instabilities. 
Because for spin-unrestricted calculations the $\hat{S}_z$-spin symmetry is imposed (i.e., each spin-orbital has a definite spin $\uparrow$ or $\downarrow$), the spin-flipped excitations (when the spin-orbitals $i$ and $a$ have different spins, or the spin-orbitals $j$ and $b$ have different spins) do not contribute to the dRPA-I, dRPA-II, RPAx-I, SOSEX, and RPAx-SO2 correlation energy variants. This can be exploited to reduce the cost of the calculations and to avoid that the evaluation of the correlation energy in these RPA variants be contaminated by spin-flipped instabilities. On the other hand, the spin-flipped excitations do contribute to the RPAx-II correlation energy variant~\cite{KloTeaCorPedHel-CPL-11} which is thus subject to spin-flipped instabilities. For this reason, and because of its general poor performance~\cite{Hes-JCP-11,TouZhuSavJanAng-JCP-11,AngLiuTouJan-JCTC-11}, we do not consider the RPAx-II variant in this work.

RPA calculations can be carried out with and without range separation.
For the full-range case, RPA correlation energies are evaluated
using orbitals and orbital energies from a HF self-consistent calculation.
For the range-separated case, the self-consistent-field (SCF) starting point is a range-separated hybrid (RSH) calculation~\cite{Angyan:05}:
\equ{eq:RSH}{
E_\text{RSH}=
    \min_{\Phi}\left\{
       \bra{\Phi} \hat{T}+\hat{V}_{\text{ne}}+\hat{W}_{\text{ee}}^\text{\text{lr}} \ket{\Phi}
      +E_{\text{Hxc}}^\text{\text{sr}} \left[ n_{\Phi} \right]
               \right\}
,}
where $\Phi$ is a single-determinant wave function of density $n_\Phi$, $\hat{T}$ is the kinetic energy operator, $\hat{V}_{\text{ne}}$ is the nuclei-electron interaction operator, $\hat{W}_{\text{ee}}^\text{lr}$ is a long-range electron-electron interaction operator associated with the long-range interaction $w_{\text{ee}}^\text{lr}(r)=\text{erf}(\mu r)/r$, and $E_{\text{Hxc}}^\text{sr}[n]$ is the corresponding $\mu$-dependent short-range Hartree-exchange-correlation density functional. The parameter $\mu$ controls the range of the separation. The RSH energy does not contain long-range correlation, which is added afterwards by evaluating the RPA correlation energies using long-range two-electron integrals and RSH orbitals and orbital energies.

\section{Computational Details}
\label{sec:compdetails}

All calculations have been performed with a development version of the {\tt MOLPRO} 2012 program~\cite{MOLPRO:12},
in which our previous spin-restricted implementations of the RPA correlation energy expressions have been generalized to a spin-unrestricted formalism. We use a straightforward implementation in which all the RPA methods scale as $N_\text{o}^3 N_\text{v}^3$ where $N_\text{o}$ and $N_\text{v}$ are the numbers of occupied and virtual orbitals.

The full-range calculations (post-HF calculations) are labelled HF+dRPA-I, HF+dRPA-II, etc. 
For the range-separated calculations, we use the
short-range Perdew-Burke-Ernzerhof (srPBE) exchange-correlation functional of Ref.~\onlinecite{GolWerStoLeiGorSav-CP-06}
(which is a modified version of the one of Ref.~\onlinecite{Toulouse:05a}), and the post-RSH calculations are labelled RSH+dRPA-I, RSH+dRPA-II, etc.
The effect of the choice of the short-range density functional was studied in a number of previous works (see, e.g., Refs.~
\onlinecite{ZhuTouSavAng-JCP-10,IreHenScu-JCP-11,GolErnMoeSto-JCP-09}
).
We do not expect a large dependence of the results on the short-range density functional, nevertheless a comprehensive study on the subject would be useful.

\figCOMPARE

We have performed a wide range of RPA calculations on the AE6 and BH6 datasets~\cite{LynTru-JPCA-03} to assess the dependence on the basis and on the range-separation parameter $\mu$, as well as the performance of the different methods. The AE6 dataset is a small representative benchmark of six atomization energies consisting of SiH$_4$, S$_2$, SiO, C$_3$H$_4$ (propyne), C$_2$H$_2$O$_2$ (glyoxal), and C$_4$H$_8$ (cyclobutane). The BH6 dataset is a small representative benchmark of forward and reverse hydrogen transfer barrier heights of three reactions, OH + CH$_4$ $\rightarrow$ CH$_3$ + H$_2$O, H + OH $\rightarrow$ O + H$_2$, and H + H$_2$S $\rightarrow$ HS + H$_2$. We compute mean absolute deviations (MADs) as a function of the range-separation parameter $\mu$ and as a function of the cardinal number of Dunning cc-pVXZ basis sets~\cite{Dunning:89}. All the calculations for the AE6 and BH6 datasets were performed at the geometries optimized by quadratic configuration interaction with single and double excitations with the modified Gaussian-3 basis set (QCISD/MG3)~\cite{AE6BH6web}. The reference values for the atomization energies and barrier heights are the non-relativistic FC-CCSD(T)/cc-pVQZ-F12 values of Refs.~\onlinecite{Haunschild:12a,Haunschild:13}.

We have further tested the MP2, dRPA-I, and RPAx-SO2 methods on a set of 49 atomization energies~\cite{Fast:99} referred here to as the AE49 dataset (consisting of the G2-1 dataset~\cite{Curtiss:91,Curtiss:97} stripped of the six molecules containing Li, Be, and Na) and on the DBH24/08 dataset~\cite{Zheng:07,Zheng:09} of 24 forward and reverse reaction barrier heights. These calculations were performed with the cc-pVQZ basis set, with MP2(full)/6-31G* geometries for the AE49 dataset, and with the aug-cc-pVQZ basis set with QCISD/MG3 geometries for the DBH24/08 dataset. We have carried out a statistical analysis of the results in the form of normal distribution of errors. The reference values for the AE49 dataset are the non-relativistic FC-CCSD(T)/cc-pVQZ-F12 values of Ref.~\onlinecite{Haunschild:12b}, and the reference values for the DBH24/08 dataset are the zero-point exclusive values from Ref.~\onlinecite{Zheng:09}.

The calculations labelled dRPA-I, dRPA-II and RPAx-I correspond to Eqs.~(\ref{eq:ACdRPAI})-(\ref{eq:ACRPAxI}), where the integral over the coupling constant is carried out by a 7-point Gauss-Legendre quadrature. The calculations designated by SOSEX and RPAx-SO2 correspond to Eqs.~(\ref{eq:rCCDSOSEX}) and~(\ref{eq:rCCDSO2}). The Riccati equations are solved iteratively, as described in Ref.~\onlinecite{TouZhuSavJanAng-JCP-11}. Core electrons are kept frozen in all our calculations. Spin-restricted calculations are performed for all the closed-shell systems, and spin-unrestricted calculations for all the open-shell systems. For a few systems, the full-range RPAx-I integrand in Eq.~(\ref{eq:ACRPAxI}) diverges in the vicinity of $\alpha=1$~\cite{KloTeaCorPedHel-CPL-11}, but in practice in our calculations since the quadrature abscissae do not exceed $\alpha=0.975$ this divergence is avoided. A more detailed study of instabilities will be shown elsewhere.
In practice, we did not encounter instabilities for the full-range HF+RPAx-SO2 method, provided of course that the true SCF minimum was found. Furthermore, we did not observed any instabilities for the range-separated RSH+RPAx-I and RSH+RPAx-SO2 methods.

\section{Results and discussion}
\label{sec:results}

Throughout the paper, the following color and point symbol code is used: MP2 results are shown in black and their symbol is a circle; dRPA-I figures are displayed in blue, their symbol is a square; and RPAx-SO2 data are drawn in red, their symbol being a triangle. When a distinction needs to be done between full-range calculations and range-separated calculations, the former are empty symbols and the latter are solid symbols.

Figure~\FloatRef{fig:depMU} shows MADs for the AE6 and BH6 datasets as a function of the range-separation parameter $\mu$, from $\mu=0.3$ bohr$^{-1}$, where all methods nearly coincide since the long-range correlation contribution is small, to full-range calculations ($\mu=\infty$), where the methods yield very different results. Range separation greatly reduces the MADs on the two datasets for all the methods. For the AE6 dataset, the usually used value of $\mu=0.5$~\cite{GerAng-CPL-05a} for the range-separation parameter yields results reasonably close to optimal for all the methods, with MADs around 5 kcal/mol. In the case of the BH6 dataset, although the value of $\mu=0.6$ gives overall the best results, the MADs obtained with $\mu=0.5$ lie within 1 kcal/mol of the minimal MADs. Consequently, we choose to use the value of $\mu=0.5$ in all the following. Note that for this value of $\mu$ all the range-separated methods give about the same MADs.

\tabMUE
\figNDist

Figure~\FloatRef{fig:depBASIS} shows the MADs for the AE6 and BH6 datasets as a function of the cardinal number X of the Dunning cc-pVXZ basis sets (X=2,3,4). For the AE6 dataset, the overall gain in MAD when going from cc-pVDZ to cc-pVQZ calculations is around 25 kcal/mol for all full-range methods except for RPAx-SO2 whose MAD increases (the mean deviation changes sign, not shown), while the gain for range-separated calculations is around 10 kcal/mol. Similarly, the gain when going from cc-pVTZ to cc-pVQZ is about 5 kcal/mol for full-range calculations and is negligible for range-separated calculations. The full-range calculations are thus not yet converged with the cc-pVQZ basis set while the range-separated calculations can be considered as converged with the same basis set. This clearly demonstrates that range-separated methods are much less basis-set dependent than full-range methods for calculations of atomization energies. These observations were expected given the demonstrated exponential convergence of long-range correlation energies with respect to the cardinal number~\cite{FraMusLupTou-JJJ-XX}. For reaction barrier heights, both the full-range and range-separated methods are relatively weakly dependent on the basis set.

\figMeanVar

We now discuss results for the full-range and range-separated MP2, dRPA-I, and RPAx-SO2 methods on the larger AE49 and DBH24/08 datasets. Table~\FloatRef{tab:MAD} shows MADs for selected values of the range-separation parameter $\mu$. For the full-range case ($\mu=\infty$), the MADs that we obtain on the AE49 dataset for HF+MP2 and HF+dRPA-I (5.63 kcal/mol and 23.23 kcal/mol, respectively) are in good agreement with the MADs on the G2-I dataset reported with the same methods with a different basis set in Ref.~\onlinecite{RenRinBluWieTkaSanReuSch-NJP-12} (5.9 kcal/mol and 21.7 kcal/mol, respectively).

For range-separated calculations, the results confirm that, among the values tested, $\mu=0.5$ yields the smallest MADs for all the methods for the two datasets. For this value of $\mu$, the RSH+RPAx-SO2 method gives the smaller MAD of 4.06 kcal/mol for the AE49 dataset, slightly better than RSH+MP2 which gives a MAD of 5.09 kcal/mol. For the DBH24/08 dataset, at $\mu=0.5$, the RSH+MP2, RSH+dRPA-I, and RSH+RPAx-SO2 methods all give similar MADs of about 3 kcal/mol. These results confirm the representativeness of the AE6 and BH6 datasets and justifies the subsequent analysis of the range-separated MP2, dRPA-I, and RPAx-SO2 methods at $\mu=0.5$.

We note that full-range RPA calculations are often performed as post-Kohn-Sham calculations, i.e. using Kohn-Sham orbitals and orbital energies. We choose instead to perform full-range RPA calculations as post-HF calculations because it corresponds to the limit of RSH+RPA calculations for $\mu\to\infty$. It turns out that the MAD that we obtain with HF+RPAx-SO2 on AE49 (12.20 kcal/mol) is similar to the MADs reported for post-PBE dRPA-I calculations on the G2-I dataset (13.3 kcal/mol~
\cite{RenRinBluWieTkaSanReuSch-NJP-12}
 or 10.2 kcal/mol~
\cite{PaiRenRinScuGruKreSch-NJP-12}
 depending on the basis sets).

The detailed results of the calculations on the AE49 and DBH24/08 datasets for the full-range and range-separated MP2, dRPA-I, and RPAx-SO2 methods at $\mu=0.5$ are given in the supplementary material~\cite{MusReiAngTou-JJJ-XX-note}. These data are analyzed in Figures~\FloatRef{fig:NDistAE55} and~\FloatRef{fig:NDistBH24} using distributions of errors. The bins correspond to the actual distributions of the errors and the curves are fitted Gaussian distributions of mean $m$ and standard deviation $\sigma$. For the AE49 dataset, full-range MP2 gives a small mean error of $m=-0.57$ kcal/mol but with a standard deviation of $\sigma=7.02$ kcal/mol, while full-range dRPA-I strongly underestimates (with a mean of $m=-23.23$ kcal/mol) and full-range RPAx-SO2 overestimates ($m=9.33$ kcal/mol) the atomization energies. When going from full-range to range-separated calculations, the three methods give much narrower distributions of the errors, and for the case of dRPA-I and RPAx-SO2 much smaller mean errors of $m=-6.43$ and $m=-2.90$ kcal/mol, respectively.

For the DBH24/08 dataset, the three full-range methods tend to overestimate the reaction barrier heights. Range separation does not change much the standard deviations (which range from 4.81 kcal/mol to 7.08 kcal/mol), but it greatly reduces the mean errors which are $m=0.20$, $m=0.55$, and $m=-0.19$ kcal/mol for RSH+MP2, RSH+dRPA-I, and RSH+RPAx-SO2, respectively.

Finally, the means $m$ and standard deviations $\sigma$ of the fitted Gaussian distributions are reported in Figure~\FloatRef{fig:MeanVar}, which gives a way of easily assessing the performance of the full-range and range-separated methods by their position in the ($m,\sigma$) plane. If we define the best method as the one closest to the point (0,0) in this plane, the best method is RSH+RPAx-SO2 for atomization energies while RSH+MP2 and RSH+RPAx-SO2 are about equally good for reaction barrier heights.

Given that for atomization energies and reaction barrier heights
RSH+MP2 is about as accurate as RSH+RPAx-SO2 while being less expensive,
one could conclude that RSH+MP2 is a better method.
However, RSH+RPAx-SO2 is significantly more accurate that RSH+MP2 for weak intermolecular interactions
(see Ref.~[
\onlinecite{TouZhuSavJanAng-JCP-11}
]),
hence RSH+RPAx-SO2 appears as a more systematic method for various types of applications.

\section{Conclusion}
\label{sec:conclusion}

We have implemented several RPA correlation energy variants in a spin-unrestricted formalism with or without range separation, and tested them on thermochemistry datasets of atomization energies and reaction barrier heights. Range-separation greatly improves the accuracy of all the RPA variants. Specifically, the RSH+RPAx-SO2 variant is among the most accurate range-separated RPA variant with mean absolute deviations of about 4 kcal/mol for atomization energies and about 3 kcal/mol on reaction barrier heights. Since RSH+RPAx-SO2 had already been shown to be among the most accurate range-separated RPA variant for weak intermolecular interactions~\cite{TouZhuSavJanAng-JCP-11}, this work confirms RSH+RPAx-SO2 as a promising method for general chemical applications.

\section*{Acknowledgements}
We thank Wuming Zhu (Hangzhou, China) for discussions in the early stage of this work.

\bibliographystyle{jchemphys}

\end{document}